\documentstyle[12pt]{article}

\begin{document}
\title{Broken unitarity of the SM and a new theory of EW Interactions
\footnote{Talk presented at Intern. Symp. on Frontiers of Science,
6/17-6/19/02, Beijing, China}}
\author{Bing An Li\\
Department of Physics and Astronomy, University of Kentucky\\
Lexington, KY 40506, USA}

\maketitle

\begin{abstract}
It is shown that in an axial-vector field theory the axial-vector field
is always accompanied by a spin-0 field which has negative metric. Therefore,
unitarity is broken.
The same results are found in a theory of charged vector fields which are
coupled to two fermions whose masses are different.
These results are applied to the SM. It is found that both Z and W fields
contain spin-0 component.
Their masses are \(m_{\phi^0}=m_t
e^{28.4}=3.78\times10^{14} GeV\) and \(m_{\phi^\pm}=m_t
e^{27}=9.31\times10^{13}GeV\) respectively.
They have negative
metric which leads to negative probability. Therefore, the unitarity of
the SM is broken at about $\sim 10^{14}$ GeV.
The masses of W and Z can be generated by two types of interactions mentioned 
above. A new Lagrangian has been constructed, which is the same as the one
of the SM without Higgs sector and the fermions are massive. 
The masses of the W and
the Z bosons are obtained to be \(m^{2}_{W}={1\over2}g^{2}m^{2}_{t}\)
and \(m^{2}_{Z}=\rho m^{2}_{W}/cos^{2}\theta_{W}\) with
\(\rho\simeq 1\). \(G_F={1\over 2\sqrt{2}m^2_t}\). A cut-off which is less 
than $10^{14}$ GeV has to be introduced to the new theory.
\end{abstract}

\section{Introduction}

The Lagrangian of the SM after spontaneous symmetry breaking is
\begin{eqnarray}
\lefteqn{{\cal L}=
-{1\over4}A^{i}_{\mu\nu}A^{i\mu\nu}-{1\over4}B_{\mu\nu}B^{\mu\nu}
+\bar{q}\{i\gamma\cdot\partial-m_q\}q}
\nonumber \\
&&+\bar{q}_{L}\{{g\over2}\tau_{i}
\gamma\cdot A^{i}+g'{Y\over2}\gamma\cdot B\}
q_{L}+\bar{q}_{R}g'{Y\over2}\gamma\cdot Bq_{R}\nonumber \\
&&+\bar{l}\{i\gamma\cdot\partial-m_{l}\}l
+\bar{l}_{L}\{{g\over2}
\tau_{i}\gamma\cdot A^{i}-{g'\over2}\gamma\cdot B\}
l_{L}-\bar{l}_{R}g'\gamma\cdot B l_{R}\nonumber \\
&&+{1\over2}m^2_ZZ_\mu Z^\mu+m^2_W W^+_\mu W^{-\mu}+{\cal L}_{Higgs}.
\end{eqnarray}

Comparing with QED and QCD, there are two new interactions,
taking (t,b) generation as an example, 
Z, and W are
\begin{equation}
{\cal L}={\bar{g}\over4}
\bar{t}\gamma_\mu\gamma_5  tZ^\mu
-{\bar{g}\over4}
\bar{b}\gamma_\mu\gamma_5 bZ^\mu,
\end{equation}
\[{\cal L}={g\over2\sqrt{2}}\bar{t}\gamma_\mu(1+\gamma_5)b W^{+\mu}+
+\bar{b}\gamma_\mu(1+\gamma_5)t W^{-\mu}.\]
\[m_t\neq m_b\]
It has been known for a long time that the unitarity of renormalizability of the SM
after spontaneous symmetry breaking has been proved in Ref.[1]. However,
these two new vertices are not included in 't Hooft's paper[1].
We need to study their effects. 
\section{Theory of axial-vector field}
The Lagrangian of a model of an axial-vector field and a
fermion is constructed as[2]
\begin{equation}
{\cal L}=-{1\over4}(\partial_\mu a_\nu-\partial_\nu a_\mu)^2
+\bar{\psi}\{i\gamma\cdot
\partial+e\gamma\cdot a\gamma_{5}\}\psi-m\bar{\psi}\psi.
\end{equation}
The amplitude of the vacuum polarization of axial-vector field
is
\begin{equation}
\Pi^a_{\mu\nu}={1\over2}(p_\mu p_\nu-p^2 g_{\mu\nu})F_{a1}(z)+F_{a2}(z)
p_\mu p_\nu+{1\over2}m^2_a
g_{\mu\nu},
\end{equation}
where
\[
F_{a1}(z)=1+
\frac{e^2}{(4\pi)^2}[{1\over3}D\Gamma(2-{D\over2})({\mu^2\over m^2})
^{\epsilon\over2}-8f_1(z)+8f_2(z)]\}\}\]
\[f_2(z)={1\over z}\int^1_0 dx log\{1-x(1-x)z\}, \]
\[F_{a2}(z)=-\frac{4e^2}{(4\pi^2)}f_2(z), \]
\[m^2_a=
\frac{2e^2}{(4\pi^2)}D\Gamma(2-{D\over2})({\mu^2\over
m^2})^{\epsilon\over2}m^2.\]

The function $F_{a1}$ is used to renormalize the $a_\mu$ field.
A mass term is generated by massive fermion loop. $F_{a2}$
is another new term
generated by the axial-vector 
coupling between axial-vector field and massive fermion. The function
$F_{a2}$ is finite and rewritten as
\begin{equation}
F_{a2}(z)=\xi+(p^2-m^2_\phi)G_{a2}(p^2),
\end{equation}
where $m^2_\phi$ is the mass of a spin-0 state whose existence will be
studied below,
$G_{a2}$ is the radiative correction of the term $(\partial_\mu a^\mu)^2$, and
\begin{equation}
\xi=F_{a2}({m^2_\phi\over m^2}).
\end{equation}
The new perturbation theory of axial-vector field theory is constructed as
\begin{eqnarray}
\lefteqn{{\cal L}_{a0}=-{1\over4}(\partial_\mu a_\nu-\partial_\nu a_\mu)^2
+\xi(\partial_\mu a^\mu)^2+{1\over2}m^2_a a^2_\mu,}\\
&&{\cal L}_{ai}=e
\bar{\psi}\gamma_\mu\gamma_5\psi a^\mu+{\cal L}_c,\\
&&{\cal L}_c=
-\xi(\partial_\mu a^\mu)^2-{1\over2}m^2_a a^2_\mu.
\end{eqnarray}
${\cal L}_c$ is a counter term of the Lagrangian.
${\cal L}_{a0}$ is the Stueckelberg's Lagrangian. The 
axial-vector field $a_{\mu}$ has four independent components.

The equation satisfied by $\partial_\mu a^\mu$ is derived
\begin{equation}
\partial^2(\partial_\mu a^\mu)-{m^2_a\over2\xi}(\partial_\mu a^\mu)=0.
\end{equation}
$\partial_\mu a^\mu$ is a pseudoscalar field and we define
\begin{equation}
\partial_\mu a^\mu=b\phi.
\end{equation}
The equation of the new field $\phi$ is found
\begin{equation}
\partial^2\phi-{m^2_a\over2\xi}\phi=0,\;\;\;m^2_{\phi}=-{m^2_a\over2\xi}.
\end{equation}
the mass of the $\phi$ boson is the solution of the equation
\begin{equation}
2F_{a2}({m^2_\phi\over m^2})m^2_\phi+m^2_a=0.
\end{equation}
In order to show the existence of solution
we take
\[{e^2\over\pi^2}{m^2\over m^2_a}=1\]
as an example.
The numerical calculation shows that $\xi< 0$ and
the solution is found to be
\[m_\phi=8.02m.\]
If
\[{e^2\over\pi^2}{m^2\over m^2_a}=0.5\]
is taken we obtain
\[m_\phi=20.42 m.\]
The value of $m_\phi$ increases while $e^2$ decreases.
It is necessary to point out that $F_{a2}(z)$ is negative in the region of
the mass of $\phi$ boson.
We separate $a_\mu$ field into a massive spin-1 filed $a'_\mu$ and a
pseudoscalar
filed $\phi$
\begin{eqnarray}
\lefteqn{a_\mu=a'_\mu+c\partial_\mu \phi,}\\
&&\partial_\mu a'^\mu=0.
\end{eqnarray}

The free Lagrangian is divided into two parts
\begin{eqnarray}
\lefteqn{{\cal L}_{a0}={\cal L}_{a'0}+{\cal L}_{\phi 0},}\\
&&{\cal L}_{a'0}=-{1\over4}(\partial_\mu a'_\nu-\partial_\nu a'_\mu)^2
+{1\over2}m^2_a a'_\mu a'^\mu,\\
&&{\cal L}_{\phi 0}={1\over2m^2_\phi}\partial_\mu\phi\{
\partial^2+m^2_\phi\}\partial^\mu\phi.
\end{eqnarray}
The coefficient c is determined by the normalization of
${\cal L}_{\phi 0}$
\begin{equation}
c=\pm{1\over m_a},
\end{equation}
and we obtain
\begin{equation}
b=-cm^2_\phi=\mp{m^2_\phi\over m_a}.
\end{equation}
The signs don't affect the physical
results when $\phi$ appears as virtual particle.
The propagator of the $a_\mu$ field is derived
\begin{equation}
\Delta_{\mu\nu}=
\frac{1}{p^2-m^2_a}\{-g_{\mu\nu}+(1+{1\over2\xi})
\frac{p_\mu p_\nu}{p^2-m^2_\phi}\}.
\end{equation}
It can be rewritten as
\begin{equation}
\Delta_{\mu\nu}=\frac{1}{p^2-m^2_a}\{-g_{\mu\nu}+\frac{p_\mu p_\nu}
{m^2_a}\}-{1\over m^2_a}\frac{p_\mu p_\nu}{p^2-m^2_\phi}.
\end{equation}
The first part is the propagator of the massive spin-1 field and
the second part is the propagator of the pseudoscalar field.
The propagator of the pseudoscalar is determined to be
\begin{equation}
-\frac{1}{p^2-m^2_\phi}.
\end{equation}
It is different from the propagator of a regular spin-0 field by a minus sign.
There are
problems of indefinite metric and negative probability when the $\phi$ field
is on mass shell. $\phi$ is a ghost.
Therefore, in the energy region of $m_\phi$ the unitarity
of this theory is broken.
\section{Theory of charged vector fields}
In this section we study a theory of charged vector
fields which
are coupled to two fermions whose masses are different.
The Lagrangian is constructed as[2]
\[{\cal L}=-{1\over2}(\partial_\mu v^{+}_\nu-\partial_\nu v^{-}_\mu)^2
+e(\bar{u}\gamma_\mu dv^{-}_\mu+\bar{d}\gamma_\mu uv^{+}_\mu)
-m_u\bar{u}u-m_d\bar{d}d.\]
The amplitude of the vacuum polarization is derived
\[\Pi^{v^*}_{\mu\nu}={1\over2}(p_\mu p_\nu-p^2 g_{\mu\nu})F_{v^* 1}(p^2)
+F_{v^* 2}(p^2)p_\mu p_\nu+{1\over2}m^2_{v^*}
g_{\mu\nu},\]
where
\[m^2_{v^*}=\frac{4D}{(4\pi)^2}g^2\Gamma(2-{D\over 2})\int^1_0
dx({\mu^2\over L_0})^{\epsilon\over2}m_-\{m_- +m_{+}(2x-1)\},\]
\[F_{v^* 1}(p^2)=\frac{4D}{(4\pi)^2}g^2
\Gamma(2-{D\over 2})
\int^1_0 dx x(1-x)
({\mu^2\over L})^{\epsilon\over2},\]
\[F_{v^* 2}(p^2)=-\frac{4D}{(4\pi)^2}g^2{1\over p^2}\int^1_0 dx m_-
\{m_- +m_+ (2x-1)\}log\{1-{1\over L_0}x(1-x)p^2\},\]
where
\[m_- ={1\over2}(m_u-m_d),\;\;\;m_+ ={1\over2}(m_u+m_d),\]
\[L_0=xm^2_u+(1-x)m^2_d,\;\;\; L=L_0-x(1-x)p^2.\]
The equations show that
nonzero $m^2_{v^*}$ and $F_{v^* 2}$ are resulted in \(m_u\neq m_d\).

The structure of the vacuum polarization is the same as the case of 
axial-vector field.
Both mass and divergence of the vector fields are generated by massive 
fermion loop.
Therefore,
the charged vector fields have
four independent components too: three spin-1 components and one scalar
whose mass can be determined.
The scalar
component has negative metric which leads to negative probability. The unitarity
of this theory is broken at the mass of the scalar.
\section{Weinberg's $2^nd$ sum rule}
Pion, $\rho$ meson, and $a_1$ meson are made of u and d quarks.\\
Why pion is so light ?\\
Why $\rho$ meson is much heavier than pion?\\
Why $a_1$ meson is much heavier than $\rho$ meson?\\
It is well known that pion mass is
originated in explicit chiral symmetry breaking, the Gell-Mann, Oakes,
and Ranner formula. In 1967 Weinberg published a paper[3]
in which the Weinberg's second sum rule is
obtained
\[m^2_a=2m^2_\rho.\]
with an assumption about the high
energy behavior of the propagator of the axial-vector fields.

The question is that based on QCD how can we understand the masses of pion,
$\rho$, $a_1$, and Weinberg's rule. 
Masses of mesons are associated with some kind of symmetry breaking.
In QCD there are only two mechanism of breaking chiral symmetry:
explicit and dynamical chiral symmetry breaking. \\
{\bf Where is the third one?}
In Ref.[4] we have proposed an effective chiral theory of light mesons.
Taking two flavors as an example,
\begin{eqnarray}
{\cal L}=\bar{\psi}(x)(i\gamma\cdot\partial+\gamma\cdot v
+\gamma\cdot a\gamma_{5}
-mu(x))\psi(x)-\bar{\psi(x)}M\psi(x)\nonumber \\
+{1\over 2}m^{2}_{0}(\rho^{\mu}_{i}\rho_{\mu i}+
\omega^{\mu}\omega_{\mu}+a^{\mu}_{i}a_{\mu i}+f^{\mu}f_{\mu})
\end{eqnarray}
where \(a_{\mu}=\tau_{i}a^{i}_{\mu}+f_{\mu}\), \(v_{\mu}=\tau_{i}
\rho^{i}_{\mu}+\omega_{\mu}\),
and \(u=exp\{i\gamma_{5}(\tau_{i}\pi_{i}+
\eta)\}\), and M is the current quark matrix.
Since mesons are bound states solutions of $QCD$ they
are not independent degrees of freedom. Therefore,
there are no kinetic terms for meson fields. The kinetic terms
of meson fields are generated from quark loops.
m is a parameter, the constituent quark mass, which related to quark
condensate, dynamical chiral symmetry breaking.\\
This theory has following features:
\begin{enumerate}
\item The theory is chiral symmetric in the limit of $m_{q}\rightarrow 0$,
\item The constituent quark mass is introduced as m and
the theory has dynamically chiral symmetry breaking(m),
\item VMD is a natural result
\[{e\over f_{v}}\{-{1\over2}F^{\mu\nu}(\partial_{\mu}\rho_{\nu}-
\partial_{\nu}\rho_{\mu})+A^{\mu}j^{\mu}\}.\]
\item Axial-vector currents are bosonized
\[-{g_{W}\over4f_{a}}{1\over f_{a}}\{-{1\over2}F^{i\mu\nu}(\partial_{\mu}
a^{i}_{\nu}-\partial_{\nu}a^{i}_{\mu})+A^{i\mu}j^{i}_{\mu}\}\\
-{g_{W}\over4}\Delta m^{2}f_{a}A^{i}_{\mu}a^{i\mu}\\
-{g_{W}\over4}f_{\pi}A^{i\mu}\partial_{\mu}\pi^{i},\]
\item The Wess-Zumino-Witten anomalous action is the leading term of the
imaginary part of the effective Lagrangian,
\item Weinberg's first sum rule is satisfied analytically,
\item All the 10 coefficients of the ChPT are predicted by this theory.
ChPT is the low energy limit of the effective chiral theory of mesons.
\item The theory is phenomenologically successful:
theoretical results of the masses and strong, E$\&$M,
and weak decay widths of mesons agree well with data,
\item The form factors of pion, ${\pi}_{l3}$, $K_{l3}$, $\pi\rightarrow
e\gamma\nu$, and K$\rightarrow e\gamma\nu$ are obtained and agree with data,
\[<r^{2}>_{\pi}=0.445fm^{2},\;\;Exp.=0.44\pm0.01fm^{2},\;\;\rho-pole=
0.39fm^{2}.\]
$\pi\pi$ and $\pi$K scatterings are studied. Theory agrees with data,
\item The parameters of this theory are: m(quark condensate), g(universal
coupling constant), and three current quark masses,
\item Large $N_{C}$ expansion is natural in this theory. All loop diagrams
of mesons are at higher orders in $N_{C}$ expansion,
\item A cut-off has been determined to be 1.8GeV. All the masses of mesons
are below the cut-off. The theory is self consistent,
\end{enumerate}

To the leading order in quark mass expansion,
the masses of the octet pseudoscalar and vector mesons are derived
\begin{eqnarray}
m^{2}_{\pi}=-{2\over f^{2}_{\pi}}(m_{u}+m_{d})<0|\bar{\psi}\psi|0>,
\nonumber \\
m^{2}_{\rho}=6m^2.
\end{eqnarray}
Therefore, in this theory explicit chiral symmetry breaking is responsible
for $m_\pi$ and $m_\rho$ is revealed from dynamical chiral symmetry breaking.
A new mechanism of chiral symmetry breaking is found in this theory, which
is obtained from
\[\bar{\psi}\gamma_\mu\gamma_5\tau^i\psi a^i_\mu.\]
It is not anomaly,
there are two $\gamma_5$. However, because of the anticommute property
of $\gamma_5$ an additional mass term is obtained from quark loop
\[(1-{1\over2\pi^2 g^2})m^2_a=m^2_\rho+m^2_\rho\]
On the LHS there is a factor
\[1-{1\over2\pi^2 g^2}\]
which is obtained from the the behavior of the propagator $a_1$ field at
high energy.\\
Besides the kinetic term of $a_1$ field
\[-{1\over4}(\partial_\mu a_\nu-\partial_\nu a_\mu)^2\]
there is additional divergent term
\[{1\over4\pi^2 f^2_a}(\partial_\mu a^{i\mu})^2\]
is obtained. 
The new chiral symmetry breaking originates in the axial-vector coupling
between axial-vector fields and massive fermions(m).

\section{Spin-0 component of the Z-field}
We choose unitary gauge to study the properties of Z and W fields of the SM[2]. 
Of course,
in the SM besides the diagrams of vacuum polarization by fermions there are other
diagrams in which propagators of the intermediate bosons are involved. 
However, only after taking the vacuum polarization of fermions into account
the propagator of boson can be defined. Therefore, we study the effects of
vacuum polarization of fermions first.
Using the unitary gauge, after spontaneous symmetry breaking
the free Lagrangian of W and Z bosons
in the original perturbation theory is defined as
\[{\cal L}_0=-{1\over4}(\partial_\mu Z\nu-\partial_\nu Z_\mu)^2+{1\over2}
m^2_Z Z_\mu Z^\mu\]
\[-{1\over2}(\partial_\mu W^+_\nu-\partial_\nu W^+_\mu)
(\partial^\mu W^{-\nu}-\partial^\nu W^{-\mu})+m^2_W W^+_\mu W^{-\mu}.\]
There are other
free Lagrangian
of photon, fermions and Higgs respectively.
Both W and Z are massive spin-1
fields which have three independent degrees of freedom .

We take the t- and b-quark generation as an example 
\begin{equation}
{\cal L}={\bar{g}\over4}\{
(1-{8\over3}\alpha)\bar{t}\gamma_\mu t+
\bar{t}\gamma_\mu\gamma_5  t\}Z^\mu
-{\bar{g}\over4}\{(1-{4\over3}\alpha)\bar{b}\gamma_\mu b
+\bar{b}\gamma_\mu\gamma_5 b\}Z^\mu,
\end{equation}
where \(\alpha=sin^2 \theta_W \).
The S-matrix element of the vacuum polarization of t and b quark generation
at the second order
is obtained
\begin{eqnarray}
\lefteqn{<Z|s^{(2)}|Z>=i(2\pi)^4\delta(p-p')\epsilon^\mu
\epsilon^\nu
{\bar{g}^2\over8}
\frac{N_C}{(4\pi)^2}D\Gamma(2-{D\over2})}\nonumber\\
&&\int^1_0 dx\{
x(1-x)(p_\mu p_\nu-p^2 g_{\mu\nu})[
({\mu^2\over L_t})^{{\epsilon\over2}}
[(1-{8\over3}\alpha)^2+1]\nonumber \\
&&+
({\mu^2\over L_b})^{{\epsilon\over2}}
[(1-{4\over3}\alpha)^2+1]]\nonumber \\
&&+m^2_t
({\mu^2\over L_t})^{{\epsilon\over2}}g_{\mu\nu}+m^2_b
({\mu^2\over L_b})^{{\epsilon\over2}}g_{\mu\nu}\},
\end{eqnarray}
where
\(L_t=m^2_t-x(1-x)p^2\),
\(L_b=m^2_b-x(1-x)p^2\) and \(\bar{g}^2=g^2+g'^2\).
The kinetic term is generated by both the vector and
the axial-vector couplings and the mass terms originate in the axial-vector
coupling only.

The interaction Lagrangian between Z-boson and the leptons of e and $\nu_e$ is
obtained 
\begin{equation}
{\cal L}={\bar{g}\over4}
\bar{\nu_e}\gamma_\mu(1+\gamma_5) \nu_e
Z^\mu
-{\bar{g}\over4}\{(1-4\alpha)\bar{e}\gamma_\mu e
+\bar{e}\gamma_\mu\gamma_5 e\}Z^\mu.
\end{equation}
We obtain
\begin{eqnarray}
\lefteqn{<Z|s^{(2)}|Z>=i(2\pi)^4\delta(p-p'){\bar{g}^2\over8}\epsilon^\mu
\epsilon^\nu\frac{1}{(4\pi)^2}D\Gamma(2-{D\over2})}\nonumber \\
&&\int^1_0 dx\{
x(1-x)(p_\mu p_\nu-p^2 g_{\mu\nu})[
({\mu^2\over L_e})^{{\epsilon\over2}}
[(1-4\alpha)^2+1]+
2({\mu^2\over L_\nu})^{{\epsilon\over2}}]\nonumber      \\
&&+m^2_e
({\mu^2\over L_e})^{{\epsilon\over2}}g_{\mu\nu}+m^2_\nu
({\mu^2\over L_\nu})^{{\epsilon\over2}}g_{\mu\nu}\},
\end{eqnarray}
There are other two lepton generations contributing to the vacuum
polarization.

The amplitude of the vacuum polarization of fermions is expressed as
\begin{equation}
\Pi^Z_{\mu\nu}={1\over2}F_{Z1}(z)(p_\mu p_\nu-p^2 g_{\mu\nu})+F_{Z2}(z)
p_\mu p_\nu+{1\over2}\Delta m^2_Z g_{\mu\nu},
\end{equation}
\begin{eqnarray}
\lefteqn{F_{Z1}=1+\frac{\bar{g}^2}{64\pi^2}\{\frac{D}{12}\Gamma(2-{D\over2})
[N_C y_q\sum_q(\frac{\mu^2}{m^2_q})^{{\epsilon\over2}}
+y_l\sum_l(\frac{\mu^2}{m^2_l})^{{\epsilon\over2}}]}\nonumber \\
&&-2[N_C y_q\sum_q f_1(z_q)+y_l\sum_lf_1(z_l)]+2[\sum_q f_2(z_q)
+\sum_{l=e,\mu,\tau}
f_2(z_l)]\},
\nonumber \\
&&F_{Z2}=-\frac{\bar{g}^2}{64\pi^2}\{N_C\sum_qf_2(z_q)
+\sum_{l=e,\mu,\tau}f_2(z_l)\},\\
&&\Delta m^2_Z={1\over8}{\bar{g}^2\over(4\pi)^2}D\Gamma(2-{D\over2})
\{N_c\sum_{q}m^2_q
({\mu^2\over m^2_q})^{{\epsilon\over2}}+\sum_l m^2_l({\mu^2\over m^2_l})^
{{\epsilon\over2}}\}\nonumber .
\end{eqnarray}
where \(y_q=1+(1-{8\over3}\alpha)^2\) for \(q=t,c,u\),
\(y_q=1+(1-{4\over3}\alpha)^2\)
for
\(q=b,s,d\), \(y_l=1+(1-4\alpha)^2\), for \(l=\tau, \mu, e\), \(y_l=2\) for
\(l=\nu_e,\nu_\mu,\tau_\mu\),
\(z_i={p^2\over m^2_i}\).
In the SM the Z boson gains mass from the spontaneous symmetry breaking
and the mass term $\Delta m^2_Z$
has been refereed to the renormalization of $m^2_Z$. 
Both the vector and axial-vector couplings contribute to $F_{Z1}$.
Only the axial-vector coupling contributes to $F_{Z2}$. $F_{Z2}$
is finite.

The function $F_{Z1}$ is used to renormalize the Z-field.
The term $F_{Z2}$ in Eq.(31) indicates that Z field has four independent
degrees of freedom(see section (2)).
We rewrite it as
\begin{equation}
F_{Z2}(z)=\xi_Z+(p^2-m^2_{\phi^0})G_{Z2}(p^2),
\end{equation}
where $m^2_{\phi^0}$ is the mass of a new neutral spin-0 field, $\phi^0$,
which will
be studied, $G_{Z2}$ is the radiative correction of this term, and
\begin{equation}
\xi_Z
=F_{Z2}|_{p^2=m^2_{\phi^0}}.
\end{equation}
In the original free Lagrangian Z field only has three degrees of freedom
and the divergence of Z field is zero.
Therefore,
the divergence term, $F_{Z2}$, must be included in the new "free Lagrangian" to
satisfy unitarity.
The new free Lagrangian of the Z-field
is constructed as
\begin{equation}
{\cal L}_{Z0}=-{1\over4}(\partial_\mu Z_\nu-\partial_\nu Z_\mu)^2
+\xi_Z(\partial_\mu Z^\mu)^2+{1\over2}m^2_Z Z^2_\mu.
\end{equation}
The equation of $\partial_\mu Z^\mu$ is derived 
\begin{equation}
\partial^2(\partial_\mu Z^\mu)-{m^2_Z\over2\xi_Z}(\partial_\mu Z^\mu)=0.
\end{equation}
$\partial_\mu Z^\mu$ is an independent spin-0 field.
Following section(2) we have
\begin{eqnarray}
\lefteqn{Z_\mu=Z'_\mu\pm{1\over m_Z}\partial_\mu\phi^{0},}\\
&&\partial_\mu Z'^\mu=0,\\
&&\phi^0=\mp{m_Z\over m^2_{\phi^0}}\partial_\mu Z^\mu,\\
&&\partial^2\phi^0-{m^2_Z\over2\xi_Z}\phi^0=0.
\end{eqnarray}
The mass of $\phi^0$ is obtained
\begin{eqnarray}
\lefteqn{2m^2_{\phi^0}F_{Z2}|_{p^2=m^2_{\phi^0}}+m^2_Z=0,}\\
&&m^2_{\phi^0}=-{m^2_Z\over2\xi_{Z}}.
\end{eqnarray}
\begin{equation}
3\sum_q{m^2_q\over m^2_Z}z_q f_2(z_q)+\sum_l {m^2_l\over m^2_Z}z_l
f_2(z_l)={32\pi^2\over\bar{g}^2}.
\end{equation}
For $z>4$ it is found that
\begin{equation}
f_2(z)=-{2\over z}-{1\over z}
(1-{4\over z})^{{1\over2}}log\frac{1-(1-{4\over z})^{{1\over2}}}
{1+(1-{4\over z})^{{1\over2}}}.
\end{equation}
Because of the ratios of ${m^2_q\over m^2_Z}$ and ${m^2_l\over m^2_Z}$
top quark dominates and the contributions of other fermions
can be
ignored. The equation has a solution at very large value of z. For very large z
Eq.(42) becomes
\begin{equation}
\frac{2(4\pi)^2}{\bar{g}^2}+\frac{6m^2_t}{m^2_Z}=3{m^2_t\over m^2_Z}log
{m^2_{\phi^0}\over m^2_t}.
\end{equation}
The mass of the $\phi^0$ is determined to be
\begin{equation}
m_{\phi^0}=m_t e^{\frac{m^2_z}{m^2_t}{16\pi^2\over3\bar{g}^2}+1}
=m_t e^{28.4}=3.78\times10^{14}GeV,
\end{equation}
and
\[\xi_Z=-1.18\times10^{-25}.\]
The neutral spin-0 boson is extremely heavy.

The propagator of Z boson is found 
\begin{equation}
\Delta_{\mu\nu}=
\frac{1}{p^2-m^2_Z}\{-g_{\mu\nu}+(1+\frac{1}{2\xi_Z})\frac{p_\mu p_\nu}{
p^2-m^2_{\phi^0}}\},
\end{equation}
It can be separated into two parts
\begin{equation}
\Delta_{\mu\nu}=
\frac{1}{p^2-m^2_Z}\{-g_{\mu\nu}+\frac{p_\mu p_\nu}{
m^2_Z}\}-\frac{1}{m^2_Z}\frac{p_\mu p_\nu}{p^2
-m^2_{\phi^0}}.
\end{equation}
The first part is the propagator of the physical spin-1 Z boson and
the second part is the propagator of a new
neutral spin-0 meson, $\phi^0$.

The propagator of Z boson 
takes the same form 
in the renormalization gauge in original perturbation theory. However,
from
physical point of view they are different.
The SM is gauge invariant before spontaneous symmetry breaking. Therefore,
in general, there is a gauge fixing term in the Lagrangian of the SM. In the
study presented above the gauge parameter has been chosen to be zero, unitary
gauge.
The differences between present propagator and the propagator of renormalization gauge
in the original perturbation theory of the SM are
\begin{enumerate}
\item The $\xi_z$ is determined dynamically.
As mentioned above, this result is obtained in unitary gauge.
Physics results depend on $\xi_Z$.
The gauge parameter of the renormalization gauge is determined by choosing
gauge and because of unitarity
physics results are independent of the gauge parameter.
\item In renormalization gauge ghosts are accompanied. However,
there are no additional ghosts. 
\end{enumerate}

There is a pole at \(p^2=m^2_{\phi^0}\). On the other hand,
the minus sign of Eq.(47) indicates that 
the Fock space has indefinite metric and there is problem of
negative probability when $\phi^0$ is on mass shell. 
$\phi^0$ is a ghost. Unitarity of the SM
is broken at \(E=m_{\phi^0}\).

The couplings between
$\partial_\mu \phi^0$
and
t,b,e,$\nu_e$ fermions are found
\begin{eqnarray}
\lefteqn{{\cal L}=\pm{1\over m_Z}{\bar{g}\over4}\{
(1-{8\over3}\alpha)\bar{t}\gamma_\mu t+
\bar{t}\gamma_\mu\gamma_5  t\}\partial_\mu\phi^0}\nonumber \\
&&\pm{1\over m_Z}{\bar{g}\over4}\{-(1-{4\over3}\alpha)\bar{b}\gamma_\mu b
-\bar{b}\gamma_\mu\gamma_5 b\}\partial_\mu\phi^0\nonumber \\
&&\pm{1\over m_Z}{\bar{g}\over4}\{
\bar{\nu_e}\gamma_\mu(1+\gamma_5) \nu_e
-(1-4\alpha)\bar{e}\gamma_\mu e
-\bar{e}\gamma_\mu\gamma_5 e\}\partial_\mu \phi^0.
\end{eqnarray}
The couplings with other generations of fermions take the same form.
Using the equations of fermions, it can be found there are
couplings between $\phi^0$ and fermions 
\begin{equation}
\pm{1\over m_Z}{\bar{g}\over4}2i\sum_i m_i\bar{\psi}_i\gamma_5\psi_i \phi^0,
\end{equation}
where i stands for the type of fermion. It is the same 
as the coupling between Higgs and fermion that the interaction is 
proportional to the fermion mass. However, here is pseudoscalar coupling.
\section{Spin-0 components of W-fields}
In the SM the fermion-W vertices are
\begin{equation}
{\cal L}={g\over4}\bar{\psi}\gamma_\mu(1+\gamma_5)\tau^i\psi W^{i\mu},
\end{equation}
where $\psi$ is a doublet of fermions and summation over all fermion
generations is
implicated.

The expression of the vacuum polarization of fermions is obtained[2]
\begin{equation}
\Pi^W_{\mu\nu}=F_{W1}(p^2)(p_\mu p_\nu-p^2 g_{\mu\nu})+2F_{W2}(p^2)
p_\mu p_\nu+\Delta m^2_W
g_{\mu\nu},
\end{equation}
where
\begin{eqnarray}
\lefteqn{F_{W1}(p^2)=1+{g^2\over32\pi^2}D\Gamma(2-{D\over2})\int^1_0 dx
x(1-x)\{
N_C\sum_{iq}({\mu^2\over L^i_q})^{{\epsilon\over2}}+
\sum_{il}({\mu^2\over L^i_l})^{{\epsilon\over2}}\}}\nonumber \\
&&-{g^2\over16\pi^2}\{N_C\sum_{iq} f^i_{1q}+\sum_{il}
f^i_{1l}\}+{g^2\over16\pi^2}\{N_C\sum_{iq} f^i_{2q}+\sum_{il}f_{2l}\},\\
&&F_{W2}(p^2)=-{g^2\over32\pi^2}\{N_C\sum_{iq} f^i_{2q}+\sum_{il}
f^i_{2l}\},\\
&&
\Delta m^2_W={g^2\over4}{1\over(4\pi)^2}D\Gamma(2-{D\over2})
\int^1_0 dx\{N_c\sum_{iq}L^i_q
({\mu^2\over L^i_q})^{{\epsilon\over2}}+\sum_{il}L^i_l({\mu^2\over L^i_l})^
{{\epsilon\over2}}\}.
\end{eqnarray}
where
\begin{equation}
L^1_q =m^2_b x+m^2_t (1-x),\;\;
L^2_q =m^2_s x+m^2_c (1-x),\;\;
L^3_q =m^2_d x+m^2_u (1-x),
\end{equation}
\[L^1_l =m^2_e x,\;\;
L^2_l =m^2_\mu x,\;\;
L^3_l =m^2_\tau x,\]
\begin{eqnarray}
\lefteqn{
f^i_{1q}=\int^1_0 dx x(1-x)log[1-x(1-x){p^2\over L^i_q}]},\\
&&f^i_{1l}=\int^1_0 dx x(1-x)log[1-x(1-x){p^2\over L^i_l}],\\
&&f^i_{2q}={1\over p^2}\int^1_0 dx L^i_q log[1-x(1-x)
{p^2\over L^i_q}],\\
&&f^i_{2l}={1\over p^2}\int^1_0 dx L^i_l log[1-x(1-x){p^2\over L^i_l}].
\end{eqnarray}
The function $F_{W1}(p^2)$ is used to renormalize the W-field.
In the SM W boson gains mass from spontaneous symmetry breaking and the
additional mass term has been treated by renormalization.
The divergence $F_{W2}$ leads to the existence of two charged spin-0 states,
$\phi^\pm$,
in the SM.
Now we apply the results obtained in sections(2,3) to the case of W fields.
$F_{W2}$ is rewritten as
\begin{eqnarray}
\lefteqn{F_{W2}=\xi_W+(p^2-m^2_{\phi_W})G_{W2}(p^2),}\\
&&\xi_W=F_{W2}(p^2)|_{p^2=m^2_{\phi_W}},
\end{eqnarray}
where $G_{W2}$ is the radiative correction of the term
$(\partial_\mu W^\mu)^2$ and $m^2_{\phi_W}$ is the mass of the charged spin-0
states, $\phi^\pm$, whose existence will be shown below.

The free part of the Lagrangian of W-field
is redefined as
\begin{equation}
{\cal L}_{W0}=-{1\over2}(\partial_\mu W^+\nu-\partial_\nu W^+_\mu)
(\partial_\mu W^-_\nu-\partial_\nu W^-_\mu)
+2\xi_W\partial_\mu W^{+^\mu}
\partial_\nu W^{-^\nu}\]
\[+m^2_W W^+_\mu W^{-\mu}.
\end{equation}
The equation satisfied by the divergence of the W-field
is derived
\begin{equation}
\partial^2(\partial_\mu W^{\pm\mu})-{m^2_W\over2\xi_W}(\partial_\mu
W^{\pm\mu})=0.
\end{equation}
$\partial_\mu W^{\pm\mu}$ are spin-0 fields. Therefore,
the W field of the SM has four independent
components.
The W-field is decomposed as
\begin{eqnarray}
\lefteqn{W^\pm_\mu=W'^\pm_\mu \pm{1\over m_W}\partial_\mu \phi^\pm,}\\
&&\partial\mu W'^{\pm\mu}=0,\\
&&\phi^\pm=\mp{m_W\over m^2_{\phi_W}}\partial_\mu W^{\pm\mu},
\end{eqnarray}
where $W'$ is a massive spin-1 field and $\phi^{\pm}$ are spin-0 fields.
The equation of $\phi^\pm$ is derived 
\begin{equation}
\partial^2\phi^\pm-{m^2_W\over2\xi_{W}}\phi^\pm=0.
\end{equation}

The mass of $\phi^\pm$ is determined by the equation
\begin{equation}
2m^2_{\phi_W}F_{W2}(p^2)|_{p^2=m^2_{\phi_W}}+m^2_W=0
\end{equation}
and
\begin{equation}
m^2_{\phi_W}=-{m^2_W\over 2\xi_{W}}.
\end{equation}
Numerical calculation shows that top quark is dominant in $F_{W2}$. Keeping the
contribution of top quark only, 
\begin{equation}
{p^2\over m^2_W}F_{W2}=-\frac{3g^2}{32\pi^2}{m^2_t\over m^2_W}\{-{3\over4}
+{1\over2z}
+[{1\over2}-{1\over z}+{1\over2z^2}]log(z-1)\},
\end{equation}
where \(z={p^2\over m^2_t}\).
It has a solution at very large z. At very large z
Eq.(68) becomes
\begin{equation}
{p^2\over m^2_W}F_{W2}=-{3g^2\over64\pi^2}{m^2_t\over m^2_W}logz.
\end{equation}
The mass of $\phi^\pm$ is determined to be
\begin{equation}
m_{\phi_W}=m_t e^{{16\pi^2\over3g^2}{m^2_W\over m^2_t}}=m_t e^{27}
=9.31\times10^{13}GeV,
\end{equation}
and
\[\xi_W=-3.73\times10^{-25}.\]
The charged $\phi^\pm$ are very heavy too.

The propagator of W-field is derived 
\begin{equation}
\Delta^W_{\mu\nu}=
\frac{1}{p^2-m^2_W}\{-g_{\mu\nu}+(1+\frac{1}{2\xi_W})\frac{p_\mu p_\nu}{
p^2-m^2_{\phi_W}}\},
\end{equation}
and it can be separated into two parts
\begin{equation}
\Delta^W_{\mu\nu}=
\frac{1}{p^2-m^2_W}\{-g_{\mu\nu}+\frac{p_\mu p_\nu}{
m^2_W}\}-\frac{1}{m^2_W}\frac{p_\mu p_\nu}{p^2
-m^2_{\phi_W}}.
\end{equation}
The first part is the propagator of physical
spin-1 W-field and the second
part is related to the propagator of the $\phi^\pm$ field. 
The physics of the propagator is different from
the one derived by using
renormalization gauge in the original perturbation theory
of the SM. There are no associated charged ghosts. $\xi_W$ is dynamically
generated.

The propagator shows that at very high energy there is a pole at
\(\sqrt{p^2}=m_{\phi_W}\). On the other hand, the minus sign of Eq.(73) 
indicates that there are problems of indefinite metric and negative
probability when $\phi^{\pm}$ are on mass shell. 
$\phi^{\pm}$ are ghosts. Unitarity of the SM is broken
at \(E\sim m_{\phi_W}\).

The Lagrangian of interactions between fermions and $\partial_\mu \phi^\pm$
field is
found 
\begin{equation}
{\cal L}_{q\phi}=\pm{1\over m_W}{g\over4}\sum_j\bar{\psi}_j
\gamma_\mu(1+\gamma_5)
\tau^i\psi_j
\partial^\mu\phi^{i},
\end{equation}
where j is the type of the fermion and
\begin{equation}
\phi^1={1\over\sqrt{2}}(\phi^+ +\phi^-)\;\;\;
\phi^2={1\over \sqrt{2}i}(\phi^+ -\phi^-).
\end{equation}
Using the dynamical equations of fermions of the SM, for t and b quark
generation we obtain
\begin{equation}
{\cal L}'_{q\phi}=\pm{i\over m_W}{g\over4}(m_t+m_b)
\{\bar{\psi}_t\gamma_5\psi_b\phi^+
+\bar{\psi}_b\gamma_5\psi_t\phi^-\}.
\end{equation}
It is the same as Higgs that the coupling is proportional to the fermion 
mass. However, it is pseudoscalar coupling.
\section{Effects of vacuum polarization by intermediate bosons}
Besides the vacuum polarization of fermions because of the nonlinear nature
the intermediate bosons contribute to the vacuum polarization too.
Before we proceed to study the effects of the vacuum polarization by
intermediate bosons it is necessary to restate the theoretical approach
exploited in this paper. The spin-o states of Z and W fields are revealed
from the vacuum polarization of fermions. Of course, there are vacuum
polarizations of intermediate bosons. However, the propagators of Z and W
fields can be defined only after the vacuum polarization of fermions are
taken into account. This is the reason why the vacuum polarization of
fermions has been treated differently from others.

After the propagators of Z and W are defined we can proceed to
study the effects of the vacuum polarization by intermediate bosons.
The interaction Lagrangian of intermediate bosons is obtained form the
SM
\begin{eqnarray}
\lefteqn{{\cal L}_i=i\bar{g}(\partial_\mu Z_\nu-\partial_\nu Z_\mu)
W^{-\mu}W^{+\nu}}\nonumber \\
&&+i\bar{g}Z^\mu\{(\partial_\mu W^+_\nu-\partial_\nu W^+_\mu)W^-_\nu
-(\partial_\mu W^-_\nu-\partial_\nu W^-_\mu)W^+_\nu\}\nonumber \\
&&-\bar{g}^2\{Z_\mu Z^\mu W^+_\nu W^{-\nu}-Z_\mu Z_\nu W^{+\mu}W^{-\nu}\}.
\end{eqnarray}
We calculate the contribution of W bosons to the vacuum polarization of
Z boson.
In the amplitude of the vacuum polarization of Z boson there are three
parts: kinetic term, mass term, and the term proportional to $p_\mu p_\nu$,
$F'_{Z2}p_\mu p_\nu$.
We
are interested in the last term. The
calculation shows that only the second term of the Lagrangian
contributes to $F'_{Z2}$. 
We obtain
\[F'_{Z2}=\frac{2\bar{g}^2}{(4\pi)^2}\{{1\over4}\Gamma(2-{D\over2})
({\mu^2\over m^2_W})^{{\epsilon\over2}}-{1\over12}\]
\[+{3\over2}
\int^1_0 dx[{3\over z}x(1-x)-x(5-7x)]log[1-x(1-x)z]\]
\[+\frac{2\bar{g}^2 b}{(4\pi)^2}\{-{13\over40}+{3\over2}\int^1_0 dx x^3 (1-x)\]
\[[x(m^2_{\phi_W}-m^2_W)+m^2_W][x(m^2_{\phi_W}-m^2_W)+m^2_W-x(1-x)p^2]^{-1}\]
\[-{3\over bm^2_W}\int^1_0 dxx(2x-{3\over2})[m^2_W-x(1-x)p^2]
log\{[m^2_W-x(1-x)p^2]\]
\[[x(m^2_{\phi_W}-m^2_W)+m^2_W-x(1-x)p^2]^{-1}\}\]
\[-{9\over2}{1\over p^2}\int^1_0 dx\int^x_0 dy[y(m^2_{\phi_W}-m^2_W)+m^2_W]\]
\[log\{[y(m^2_{\phi_W}-m^2_W)+m^2_W-x(1-x)p^2]\]
\[[y(m^2_{\phi_W}-m^2_W)+m^2_W]^{-1}\},\]
where \(z={p^2\over m^2_W}\) and \(b=1+{1\over2\xi_W}\).

There is divergence, therefore, renormalization of the operator
$(\partial_\mu Z^\mu)^2$ is required. $F'_{Z2}$ is written as
\begin{equation}
F'_{Z2}=F'_{Z2}(m^2_{\phi^0})+(p^2-m^2_{\phi^0})G'_{Z2}(p^2).
\end{equation}
$F'_{Z2}$ is divergent and
$G'_{Z2}$ is finite and another term of radiative correction
of $(\partial_\mu Z^\mu)^2$. We
define
\begin{eqnarray}
\lefteqn{\xi_Z+F'_{Z2}(m^2_{\phi^0})=\xi_Z Z_Z,}\\
&&Z_Z=1+{1\over \xi_Z}F'_{Z2}(m^2_{\phi^0}).
\end{eqnarray}
$Z_Z$ is the renormalization constant of the operator
$(\partial_\mu Z^\mu)^2$.
The renormalization constant $Z_Z$ defined 
guarantees the Z boson
the same propagator.

In the same way, the contribution of Z and W bosons to the vacuum
polarization of W
boson can be calculated and the renormalization constant $Z_W$ can be
defined.
After renormalization of $\partial_\mu W^{+\mu}\partial_\nu W^{-\nu}$
we still have
the same propagator of W boson.
In the same way, the contribution of Higgs to vacuum polarization can be 
studied too.
\section{\(G_F=\frac{1}{2\sqrt{2}m^2_t}\), $M_Z$, and $M_W$ without 
spontaneous symmetry breaking} 
The SM works until $10^{14}$ GeV. At this energy level unitarity in the SM 
is broken. New elements need to be added into the theory.

Let's revisit the Higgs mechanism and spontaneous symmetry breaking:
\begin{enumerate}
\item W and Z bosons gain masses
\item Unitarity of the SM. For example, $ee^+\rightarrow W^+ W^-$
\item renormalizability. For example, $WW\rightarrow ZZ$.
\end{enumerate}
However, 
\begin{enumerate}
\item Unitarity is broken at $10^{14}$ GeV and the behavior of the SM
at high energies has problem
which affects the renormalizability
\item W and Z can gain masses from fermion masses
\end{enumerate}  
We try to get rid of the Higg's mechanism and spontaneous symmetry breaking, but
keep the success of the SM. We propose to take[5]  
\begin{eqnarray}
\lefteqn{{\cal L}=
-{1\over4}A^{i}_{\mu\nu}A^{i\mu\nu}-{1\over4}B_{\mu\nu}B^{\mu\nu}
+\bar{q}\{i\gamma\cdot\partial-M\}q}
\nonumber \\
&&+\bar{q}_{L}\{{g\over2}\tau_{i}
\gamma\cdot A^{i}+g'{Y\over2}\gamma\cdot B\}
q_{L}+\bar{q}_{R}g'{Y\over2}\gamma\cdot Bq_{R}\nonumber \\
&&+\bar{l}\{i\gamma\cdot\partial-M_{f}\}l
+\bar{l}_{L}\{{g\over2}
\tau_{i}\gamma\cdot A^{i}-{g'\over2}\gamma\cdot B\}
l_{L}-\bar{l}_{R}g'\gamma\cdot B l_{R}
\end{eqnarray}
as the Lagrangian of the electroweak interactions. This is the Lagrangian of the SM
after spontaneous symmetry breaking in unitary gauge without Higgs.
$A^i_\mu$ fields are still Yang-Mills fields. The Lagrangian doesn't have
$SU(2)_L\times U(1)$ symmetry. U(1) symmetry is kept and electric current
is conserved. As mentioned above, this L has problem at $10^{14}$ GeV. A cutoff
has to be introduced under this energy. Therefore, we don't need to worry about
the renormalizability.

As shown in previous sections the masses of W and Z bosons are resulted in fermion masses
\begin{enumerate}
\item Z boson gains mass from the axial-vector couplings with massive fermions
\item W bosons gain mass from both the axial-vector couplings with massive
fermions and the vector couplings with fermions whose masses are different.
\end{enumerate}
\begin{eqnarray}
\lefteqn{{\cal L}_{M}={1\over2}\frac{N_{C}}{(4\pi)^{2}}\{{D\over4}
\Gamma(2-{D\over2})(4\pi{\mu^{2}\over m^{2}_{1}})
^{{\epsilon\over2}}+{1\over2}[1-ln(1-x)}\nonumber \\
&&-(1+{1\over x}){\sqrt{x}\over2}
ln\frac{1+\sqrt{x}}{1-\sqrt{x}}]\}m^{2}_{1}g^{2}
\sum^{2}_{i=1}A^{i}_{\mu}A^{i\mu}\nonumber \\
&&+{1\over2}\frac{N_{C}}{(4\pi)^{2}}\{{D\over4}
\Gamma(2-{D\over2})(4\pi{\mu^{2}\over m^{2}_{1}})
^{{\epsilon\over2}}-{1\over2}[ln(1-x)\nonumber \\
&&+\sqrt{x}
ln\frac{1+\sqrt{x}}{1-\sqrt{x}}]\}m^{2}_{1}(g^{2}+g'^{2})
Z_{\mu}Z^{\mu},
\end{eqnarray}
where $N_{C}$ is the number of colors and
\(x=({m^{2}_{2}\over m^{2}_{1}})^{2}\). 
It is reasonable to redefine the fermion
masses by multiplicative renormalization
\[Z_{m}m^{2}_{1}=m^{2}_{1,P},\]
\[Z_{m}=\frac{N}{(4\pi)^{2}}\{N_{G}{D\over4}\Gamma(2-{D\over2})
(4\pi)^{{\epsilon\over2}}({\mu^{2}\over m^{2}_{1}})
^{{\epsilon\over2}}+{1\over2}[1-ln(1-x)\]
\[-
(1+{1\over x}){\sqrt{x}\over2}
ln\frac{1+\sqrt{x}}{1-\sqrt{x}}]\},\]
for each generation of fermions. The index "P" is omitted in the rest
of the paper.
Now the mass of W boson is obtained
\begin{equation}
m^{2}_{W}={1\over2}g^{2}\{m^{2}_{t}+m^{2}_{b}+m^{2}_{c}+m^{2}_{s}
+m^{2}_{u}+m^{2}_{d}+m^{2}_{\nu_{e}}+m^{2}_{e}
+m^{2}_{\nu_{\mu}}+m^{2}_{\mu}+m^{2}_{\nu_{\tau}}+m^{2}_{\tau}\}
\end{equation}
Obviously, the top quark mass dominates the $m_{W}$
\begin{equation}
m_{W}={g\over\sqrt{2}}m_{t}.
\end{equation}
Using the values \(g=0.642\) and \(m_{t}=174.3\pm5.1 GeV\), it is found
\begin{equation}
m_{W}=79.1\pm2.3 GeV,
\end{equation}
which is in excellent agreement with data $80.41\pm0.056$GeV.
The Fermi coupling constant is derived 
\[G_{F}=\frac{1}{2\sqrt{2}m^{2}_{t}}=1.024(1\pm0.029)\times10^{-5}m^{-2}_{N}.\]

The mass formula of the Z boson
is written as
\begin{equation}
m^{2}_{Z}=\rho m^{2}_{W}(1+{g^{2}_{2}\over g^{2}_{1}}),
\end{equation}
where
\[\rho=(1-\frac{\alpha}{4\pi}f_4)^{-1}\]
$f_4$ is a determined quantity[5].
Ignoring the electromagnetic correction
\[\rho= 1.\]
Therefore,
\begin{equation}
m_{Z}=m_{W}/cos\theta_{W}
\end{equation}
This is the prediction of the Higgs mechanism.
\section{conclusions}
\begin{enumerate}
\item The unitarity of the SM is broken at $10^{14}$ GeV. 
A cutoff has to be introduced.
\item Based on the success of the SM a new L without Higgs sector is proposed.
\item In this new L W and Z gain masses from massive fermion loops.
Correct $m_Z$ and $m_W$ are obtained.
\item The free Lagrangian of Z and W fields are expressed by Eqs.(34,62).
\item In the new theory a cut-off which is less than $10^{14}$ GeV has
to be introduced.
\item The propagators of Z and W are
\begin{equation}
\Delta_{\mu\nu}=
\frac{1}{p^2-m^2_Z}\{-g_{\mu\nu}+(1+\frac{1}{2\xi_Z})\frac{p_\mu p_\nu}{
p^2-m^2_{\phi^0}}\},
\end{equation} 
\begin{equation}
\Delta^W_{\mu\nu}=
\frac{1}{p^2-m^2_W}\{-g_{\mu\nu}+(1+\frac{1}{2\xi_W})\frac{p_\mu p_\nu}{
p^2-m^2_{\phi_W}}\}.
\end{equation}
\item This theory predicts no Higgs. This theory can be tested by precision 
measurements. 
\end{enumerate}

This study is supported by a DOE grant.

\end{document}